\documentclass{article}
\usepackage{epsfig}
\usepackage{amsmath}
\makeatletter

\def\lsim{\compoundrel<\over\sim}
\def\compoundrel#1\over#2{\mathpalette\compoundreL{{#1}\over{#2}}}
\def\compoundreL#1#2{\compoundREL#1#2}
\def\compoundREL#1#2\over#3{\mathrel{\vcenter{\hbox{$\buildrel{#1#2}\over{#1#3}$}}}}
\begin{document}
\setcounter{page}{1}
\begin{flushright}
GUTPA/02/03/01
\end{flushright}
\renewcommand{\thefootnote}{\fnsymbol{footnote}}
\begin{center}{\LARGE{{\bf Could there be a  Fourth Generation?}}\footnote[1]{Updated and slightly extended version of an article contributed to Holger 
Bech Nielsen's Festscript.}}\end{center}
\begin{center}{\large{C.D. Froggatt}\footnote[2]{E-mail: c.froggatt@physics.gla.ac.uk} \; \large{and J.E. Dubicki}\footnote[3]{E-mail: 
j.dubicki@physics.gla.ac.uk}\\}
\end{center}
\renewcommand{\thefootnote}{\arabic{footnote}}
\begin{center}{{\it Department of Physics and Astronomy}\\{\it University of Glasgow, Glasgow G12 8QQ, Scotland}}\end{center}
\setcounter{footnote}{0}
\begin{abstract}
We investigate the possibility of incorporating a chiral fourth generation into a GUT model. We find that in order to do so, precision fits to electroweak
observables demand the introduction of light ($< M_Z$) supersymmetric particles. This also enables us to provide decay channels to the fourth-generation 
quarks. Perturbative consistency sets an upper bound on the coloured supersymmetric spectrum. The upper limit on the lightest Higgs boson is calculated and 
found to be above the present experimental lower limit.
\end{abstract}
\section{Introduction}
\label{sec1}
Despite the success of the Standard Model (SM), it is far from complete. For instance, the origin of the fermion mass hierarchy and issues 
such as baryon asymmetry are yet to be resolved. We would expect new physics to contribute to these areas. At the more fundamental level, nobody knows of 
any deep explanation as to why there should only be three generations of quarks \& leptons. For these reasons, and others, extensions of the 
three-generation SM are being investigated. Here, we investigate the possibility of incorporating a fourth-generation into a GUT model.\medskip \\ 
As is well known, all fourth-generation models must adhere to certain experimental constraints, the first of which stems from precise measurements of the 
decay characteristics of the $Z$-boson performed at LEP. This has set a lower bound of $M_F\ge\frac{M_Z}{2}$ on any non-SM particles that couple to 
the $Z$-boson. Ignoring for the moment the unnatural hierarchy emerging within the neutrino sector, we assume a Dirac mass 
$M_N \sim\left(\frac{M_Z}{2}\right)$ for the heavy neutral lepton.
We label the fourth-generation explicitly with the notation
\begin{equation}
	Q_4 = {\left( \begin{array}{c} T \\ B \end{array} \right)}_4 
			\quad T_4^c \quad B_4^c; \qquad
	L_4 = {\left( \begin{array}{c} N \\ E \end{array} \right)}_4 
			\quad N_4^c \quad E_4^c
\end{equation}
\medskip \\
Physics beyond the Standard Model is severely constrained by precision electroweak data\footnote{We assume ${\left|V_{tb}\right|}^2$, 
${\left|V_{TB}\right|}^2 \sim 1$ and ${\left|V_{Tb}\right|}^2$, ${\left|V_{tB}\right|}^2 \ll 1$ so that contributions to the
$Z \rightarrow b\overline{b}$ decay can be ignored.}.
Assuming SM contributions, fits to LEP data give the radiative correction parameter\footnote{S is the well known radiative correction parameter (weak 
isospin symmetric), normalized to zero for the SM with $M_{Higgs} =$ 100 GeV.} $S=-0.04 \pm 0.11$. For a heavy ($\gg M_Z$) degenerate fourth-generation we 
obtain $\Delta S = \frac{2}{3\pi}$ and so is ruled out at $99.8\%$ $C.L$. However, Maltoni {\it et al.}~\cite{maltoni} have shown that particles with mass 
$M \sim \frac{M_Z}{2}$ give drastically different contributions to $S$. It is possible for a heavy neutrino $N$ with mass $M_N \sim \frac{M_Z}{2}$ to 
cancel the contributions from the heavy $T,B,E$ with the SM solution\footnote{Throughout this paper, we shall employ the notation that upper case letters 
($M_x$) denote pole masses and lower case letters ($m_x$) denote running masses.}
\begin{equation}  M_E > M_N ; \quad M_T > M_B \end{equation}
\begin{equation}\label{def}  M_N \sim \frac{M_Z}{2} ; \quad M_B = M_B^{min} \end{equation}
\begin{equation}\label{fit}  
	(M_E-M_N) \sim 3(M_T-M_B) \sim 60 \; GeV
\end{equation}
where the $N$ must be relatively stable to avoid detection ({\it i.e.} mixing matrix elements $V_{Ne,\mu,\tau} < 10^{-6}$). In Eq.~(\ref{def}), $M_B^{min}$ 
is the minimum mass of the B quark as allowed by experimental searches. An extra generation can be 
accomodated below the $1\sigma$ level (or even two generations at $1.5\sigma$)\footnote{Updated fits to recent LEP data show that an extra generation 
with $M_N = M_U \sim$ 180 GeV; $M_D = M_E \sim$ 130 GeV gives a $\chi^2$ fit for the fourth-generation as good as that of the three-generation 
case~\cite{okun}. Perturbative unification is not consistent with such large masses.}.
However, it was shown by Gunion {\it et al.}~\cite{gunion} that the fourth-generation charged and neutral leptons must be relatively light 
($M_{E,N} \sim \frac{M_Z}{2}$) in order to stay in the perturbative regime below the GUT scale. Although they worked within a supersymmetric framework, 
similar results are expected to hold in the SM~\cite{pirogov}. Setting $M_E \sim \frac{M_Z}{2}$ to achieve perturbative unification is not consistent 
with the above SM fits (Eq.~(\ref{fit})) to the precision data. Therefore, we require the cancellations to arise from another sector and so we consider a 
supersymmetric theory. Indeed, it is shown in ~\cite{maltoni} that light supersymmetric particles can also effect the fit to precision data, allowing for a 
fourth-generation below the $2\sigma$ level. In particular, neutralinos (${\tilde{\chi}}^0_1$) and charginos (${\tilde{\chi}}^{\pm}_1$) with masses less 
than $M_Z$ can provide the correct sign contributions, whilst being consistent with current experimental limits if nearly degenerate~\cite{L3}\footnote{
${\tilde{\chi}}^0_1$ and ${\tilde{\chi}}^{\pm}_1$ are defined as the lightest mass eigenstates in the neutralino and chargino sectors respectively.}.
\medskip \\
In this paper we investigate the possibility of consistently incorporating a fourth-generation into a $N=1$ R-parity conserving supergravity model. 
We assume a structure akin to that of the minimal supersymmetric standard model (MSSM3), adding a complete chiral fourth-generation and its associated 
SUSY partners (the so called MSSM4). Specifically we require $(i)$ perturbative values for all Yukawa couplings at energies up to the GUT scale and 
$(ii)$ gauge coupling unification. These two constraints will be termed collectively as perturbative unification. \medskip\\
Having satisfied precision data fits, it remains for us to provide solutions that $(i)$ unify perturbatively
at the GUT scale and $(ii)$ evade the direct experimental searches performed at $CDF$. 
In section~\ref{sec2} we investigate the specific decay channels of the fourth-generation quarks and ensure we can provide consistency
with the experimental direct searches. The solution we present requires the introduction of light SUSY particles 
(${\tilde{\chi}}^{0}_1,{\tilde{\chi}}^{\pm}_1,\tilde{B}$), so that the two-body decays $T \rightarrow \tilde{B} {\tilde{\chi}}^{+}_1$
and $B \rightarrow \tilde{B} {\tilde{\chi}}^{0}_1$ are kinematically allowed. These will always dominate over the one-loop FCNC
decays $B \rightarrow b Z$ and two-generation decays $B \rightarrow c W^{-}$ that traditional searches have looked for, whilst at the 
same time suppress the decay rate $T \rightarrow b W^+$. We note a light, degenerate
chargino/neutralino pair is just what is needed to provide the necessary cancellations in precision data. \medskip \\
In section~\ref{sec3} we discuss the influence of specific fourth-generation and SUSY particle masses on precision data fits in more detail and point out 
new features that appear when trying to satisfy all constraints. In section~\ref{sec4} we proceed with a renormalisation group study of the four generation 
MSSM. Under the assumption of a common mass scale for the coloured/weak sparticle spectrum ($M_{col}$/$M_{wk}$ respectively) we will derive upper bounds 
on $M_{col}$. This bound stems from the fact that if we are to achieve perturbative values for the top quark Yukawa coupling constant $h_t$ up to the 
GUT scale, then the effects of the coloured sparticle spectrum must be included in the running of the strong coupling $\alpha_3$ at an early stage.
Finally, in section~\ref{sec5} we investigate the lightest Higgs mass in the MSSM4. \medskip \\
\section{Experiment and a Fourth Generation}
\label{sec2}
To reduce the number of free parameters we take a common mass scale for the fourth-generation quarks $M_Q$ and for the leptons $M_L$.
It is understood, however, that in reality a small non-degeneracy $\Delta M_Q = M_T - M_B$; $\Delta M_L = M_E - M_N > 0$ exists as will be emphasized in 
the following text. We begin our discussion by considering the leptonic sector. Under the assumption that the mixing between ($E,N$) and the first three 
generations is negligible, the decay $E \rightarrow N W^*$ will be dominant. Current experimental limits searching for $E \rightarrow N W^*$ from 
$e^+e^- \rightarrow E^+E^-$ production have been performed by the OPAL/L3 collaborations up to the kinematic limit $M_E \sim$ 100 GeV~\cite{lepton}. 
However, to be consistent with perturbative unification we require $M_E \sim \frac{M_Z}{2}$. Therefore, in order to evade experimental bounds, the mass 
difference $\Delta M_L$ must be less than $\sim$ 5 GeV so as to result in a virtual $W^*$ whose decay products are too soft to be triggered. Regarding the 
heavy neutrino, OPAL/L3 have set the bound $M_N >$ 70 - 90 GeV based on the search for $N \rightarrow l W^*$ ($l=e, \mu$ or $\tau$) provided the mixing 
matrix elements satisfy $V_{Ne,\mu,\tau} > 10^{-6}$~\cite{lepton}. However, we have already 
assumed that the mixing of the fourth generation leptonic sector is in fact negligible ($V_{Ne,\mu,\tau} < 10^{-6}$); so the neutrino is stable enough to 
leave the detector and, in this case, only the DELPHI bound of $M_N >$ 45 GeV from measurements of the $Z$-width applies. For the purpose of the numerical
study we take $M_L =$ 46 GeV.  \medskip \\
Turning to the quark sector, the requirement of perturbative unification places strict upper limits on the masses of the $T$ and $B$ and they certainly
must be below the top quark whose mass is $M_t = 174 \pm$ 5.1 GeV. Experimental searches for the fourth-generation quarks
have mainly concentrated on the $B$-quark where CDF search for
\begin{equation}
	e^+e^- \rightarrow B\overline{B} \rightarrow b\overline{b}ZZ \rightarrow (b\overline{b})(l^+l^-)(q\overline{q})
\end{equation}
The current bound excludes $M_B <$ 199 GeV assuming the branching ratio $BR(B \rightarrow b Z) \sim 1$. The search is also sensitive to other decay modes
as long as $BR(B \rightarrow bZ)$ is large enough to provide the leptonic trigger. For instance, the decays $B \rightarrow b h$ and 
$B \rightarrow cW^-$ will be  triggered, since the hadronic decay of the $h$ and $W$ is kinematically similar to those of the $Z$. One possible escape 
might come about if we note that $h \rightarrow N \overline{N}$ would compete 
with $h \rightarrow b \overline{b}$ for $M_h \sim M_Z$ (dominating for larger $M_h$) and would provide an invisible signature. For 
$M_h \sim M_Z$ we would also expect $BR(B \rightarrow b h) \sim BR (B \rightarrow b Z)$~\cite{hou}. However, CDF still exclude a $B$-mass 
in the range 104 GeV $\rightarrow$ 152 GeV, assuming $BR(B \rightarrow bZ) \ge 50 \%$ and no sensitivity to the other decay modes~\cite{quark}. For 
the region $M_B <$ 104 GeV, we require $BR(h \rightarrow \mbox{invisible}) >$ 30\% in order to evade the B search. This comes from the fact that
$BR(Z \rightarrow q\overline{q}) =$ 70\% and that we also assume the acceptance for the two processes
\begin{equation}
	e^+e^- \rightarrow B\overline{B} \rightarrow b\overline{b}Z\{Z \; \mbox{or} \; h\} \rightarrow (b\overline{b})(l^+l^-)\{q\overline{q}
				\; \mbox{or} \; b\overline{b} \}
\end{equation}
are equal. However, this scenario is inconsistent with the latest bounds from LEPII~\cite{invisible} which exclude $M_h <$ 106 GeV if 
$BR(h \rightarrow \mbox{invisible}) >$ 30\%. Moreover, we have yet to account for the $T$ decays 
which turn out to be highly constrained if we assume $SM$-like processes. The channel $T \rightarrow b W^+$ is prohibited for the obvious reason that 
the $T$-quark would have been picked up in the $CDF$ search for the top quark. Although we might assume that $T \rightarrow B W^*$ is dominant by 
suppressing $V_{Tb}$ we must notice that, since the $T\overline{T}$ production cross-section is similar to that of $B\overline{B}$, we would effectively 
expect double the event rate on the $B$-quark search. Taking this into account would further strengthen existing bounds on the $B$-quark mass. \medskip\\ 
From the ideas presented so far, we might conclude that the fourth-generation with perturbative unification is not consistent with experimental bounds
on the ($T,B$) quarks. We have not, however, considered the possibility of light SUSY particles providing decay channels for ($T,B$). In this situation
one can constrain the light ({\it i.e.} $< M_Z$) neutralino/chargino pair, which is already required by fits to precision data, in 
order to provide the following two-body decays 
\begin{equation}  
	T \rightarrow {\tilde{B}}_1 {\tilde{\chi}}^{\pm}_1 ; \quad B \rightarrow {\tilde{B}}_1 {\tilde{\chi}}^0_1
\label{chan}
\end{equation}
where ${\tilde{B}}_1$ is the lightest mass eigenstate, defined by the amount of mixing between the L/R weak eigenstates; 
${\tilde{B}}_1 = {\tilde{B}}_L \cos\theta_{\tilde{B}}+{\tilde{B}}_R\sin\theta_{\tilde{B}}$.
Ensuring both decay channels are kinematically accessible, combined with the fact that perturbative unification requires that  
$M_Q \le M_Q^{max} \le$ 108 GeV\footnote{$M_Q^{max}$ is defined as the maximum allowed value of $M_Q$. This is obtained by ensuring
perturbative consistency in renormalization group calculations (see section~\ref{sec4}).} (see Table~\ref{range}), places severe restrictions on the 
allowed spectrum. In the next section we combine the experimental constraints with those obtained from precision fits to obtain the allowed 
fourth-generation spectrum.
\begin{table} \hspace{3.2cm}
	\begin{tabular}{|c||c|c|c|} \hline
		$M_t$ (GeV) & 170 & 175 & 180 \\ \hline 
		$M_Q^{max}$ (GeV) & 108 & 107 & 102 \\ \hline
	\end{tabular}
\caption{Values of $M_Q^{max}$ vs.~$M_t$.}
\label{range}
\end{table}
\section{Precision Measurements and a Fourth Generation}
\label{sec3}
It is difficult to provide bounds from precision data without a fully consistent study taking into account exact particle masses, any light SUSY 
spectra present and mixings between different flavours. However, Maltoni {\it et al.}~\cite{maltoni} pointed out that a highly degenerate 
neutralino/chargino pair can provide the necessary contributions needed to cancel that of the fourth-generation, whilst at the same time being consistent 
with LEP bounds. Specifically they require
\begin{eqnarray}
	M_{\pi^+} \lsim \Delta M_{\tilde{\chi}} \lsim 3 \; GeV \qquad M_{\tilde{\chi}} & \sim & 60 \; GeV
\end{eqnarray}
where we define the notation $\Delta M_{\tilde{\chi}} =  M_{{\tilde{\chi}}^{\pm}_1} - M_{{\tilde{\chi}}^0_1}$;
$M_{\tilde{\chi}} =  M_{{\tilde{\chi}}^0_1} \sim M_{{\tilde{\chi}}^{\pm}_1}$.
Looking at their results, we see that the magnitude of the contribution to the fitted parameters from this sector are highly dependent on 
$M_{\tilde{\chi}}$. Deviations from $M_{\tilde{\chi}} \sim$ 60 GeV larger than + 30 GeV / - 5 GeV are ruled out at the $2\sigma$ level.  This 
scenario is inconsistent with perturbative unification, since $M_{\tilde{\chi}} \sim$ 60 GeV with $M_{{\tilde{B}}_1} > M_{\tilde{\chi}}$ would require 
$M_Q >$ 120 GeV in order to retain the decay channels in Eq.~(\ref{chan})\footnote{We are assuming here that the lightest supersymmetric particle
is the neutralino ${\tilde{\chi}}^0_1$.}.
One possible solution presents itself, however, if we assume that the lightest supersymmetric particle (LSP) is in fact the fourth-generation 
sneutrino $\tilde{N}$ which is stable due to R parity. Assuming the leptonic mixing angle between the third- and fourth-generation is non-zero, the 
${\tilde{\chi}}^0_1$ would decay invisibly via ${\tilde{\chi}}^0_1 \rightarrow {\upsilon}_{\tau} \tilde{N}$~\cite{gunion}. The following 
masses are possible:
\newline
\parbox{4.0cm}{\begin{eqnarray*}
		M_{\tilde{\chi}} & \simeq & 60 \; GeV \nonumber \\
		M_{{\tilde{B}}_1} & \simeq & 45 \; GeV \nonumber \\
		M_{{\tilde{N}}_1} & < & M_{{\tilde{B}}_1} \nonumber 
\end{eqnarray*}}
\hfill \parbox{8.0cm}{\begin{eqnarray}
		M_Q & = & 106 \; GeV \nonumber \\
		M_L & = & 46  \; GeV \\
		M_t & = & 170 \; GeV \nonumber
\end{eqnarray}}
\newline
where we assume that the mixing angles $\theta_{\tilde{B}/\tilde{N}}$ are such that ${\tilde{B}}_1/{\tilde{N}}_1$ 
decouple from the Z-boson at tree-level, and therefore do not contribute to its total width. The fourth-generation bottom squark will decay via 
${\tilde{B}}_1 \rightarrow c \tau {\tilde{N}}_1$. Such a decay involves the factor ${|V_{Bc} V_{N\tau}|}^2$, leading to a long lifetime. Current bounds 
looking for hadronizing sbottom  quarks exclude the range 5 GeV $\le M_{{\tilde{B}}_1} \le$ 38 GeV,  if the mixing angle $\theta_{\tilde{B}} =$ 1.17 rad 
where the production cross-section is minimized~\cite{long}. We note this corresponds to the mixing angle at which ${\tilde{B}}_1$ decouples from the 
Z-boson, as required from the Z-width constraint.  \medskip \\
Although the sparticle  masses seem to be contrived, we note that they can be obtained from reasonable assumptions about the supersymmetric 
sector. In order to obtain $M_{\pi^+} \lsim \Delta M_{\tilde{\chi}} \lsim$ 3 GeV, we require the hierarchy 
\begin{equation}
	|\mu| \gg M_1 \ge M_2
\label{hier}
\end{equation} 
where $|\mu|$ is the Higgs mixing parameter. $M_1$ and $M_2$ are the electroweak
gaugino masses (for a review of supersymmetry see~\cite{susy}). This structure can occur naturally within specific supergravity models~\cite{string}. 
The final mass eigenstates $M_{{\tilde{X}}_{1,2}}$ ($X=B,N$) depend on the appropriate squark/slepton soft mass terms and scalar tri-linear couplings that
are present in the soft supersymmetric breaking sector, and the $\mu$ term. Here, we assume it is possible to constrain these free parameters to give the 
required spectrum without contradicting the model in any way. However, more detailed studies taking into account the high-energy behaviour and 
renormalization group improvement needs to be performed, in order to establish consistency with radiative electroweak breaking and other constraints such 
as the absence of charge/colour breaking {\it etc}.\medskip \\
We note that the fits to precision data performed in~\cite{maltoni} constrained the fourth-generation masses to satisfy $M_N = M_T$ and $M_E = M_B$, 
whereas we consider the case $M_T \sim M_B > M_N \sim M_E$. From Eqs. (5.9--5.11, 5.13) in~\cite{maltoni}, we can see that as long as we satisfy the 
conditions $|M_{T,N}-M_{B,E}| \ll M_Z$, the two scenarios differ only in the universal contributions to the radiative parameters (see Maltoni, Ph.D. 
thesis~\cite{maltoni} for terminology). In particular, we would require an extra (positive) universal contribution in order to exactly reproduce the fits. 
This could arise if we were to find large SU(2) breaking within the $(\tilde{t},\tilde{b})$ sector. This has been shown to provide the correct type of 
contribution that is needed to compensate between the two scenarios~\cite{fits}.
\section{Renormalization Group Study of the MSSM4}
\label{sec4}
\begin{figure}[t]
\begin{minipage}[t]{0.75cm}
	\raisebox{3.5cm}{$M_{col}^{max}$}
\end{minipage}
\hfill
\begin{minipage}[t]{12cm}
\centerline{\epsfig{file=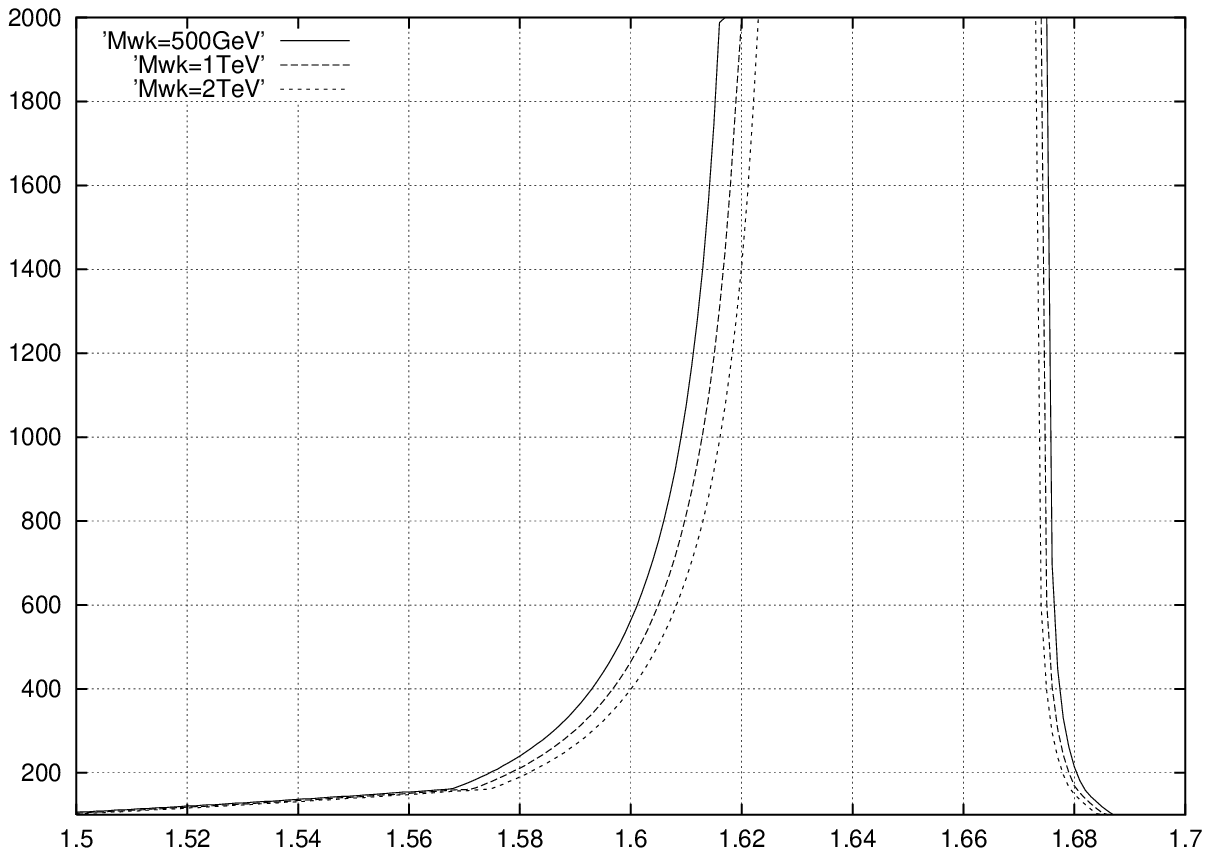,height=7cm,width=11.5cm}}
\makebox[12.5cm]{$\tan\beta$}
\end{minipage}
\caption{Plot of $M_{col}^{max}$ vs.~$\tan\beta$ for $M_Q =$ 106 GeV; $M_L =$ 46 GeV; $M_t =$ 170 GeV.}
\label{susyr}
\end{figure}

Here we investigate the effect of the fourth-generation on the evolution of couplings to the GUT scale, where we require gauge coupling unification.
This places upper limits on the masses of the extra particles to ensure their Yukawa couplings run perturbatively to the unification scale $M_U$
($h^2(\mu) \le 4\pi$, $M_Z \le \mu \le M_U$). Starting at the low-energy scale $M_Z$, the electroweak gauge couplings $\alpha_1(M_Z)$, $\alpha_2(M_Z)$
are fixed through the relations $\frac{1}{\alpha_i(M_Z)} = \frac{3}{5} \frac{\left(1-{\sin^2 \theta_W}\right)}{\alpha_{em}(M_Z)} ;
\frac{{\sin^2 \theta_W}}{\alpha_{em}(M_Z)}$
for $i=1,2$ respectively. The strong coupling $\alpha_3(M_Z)$ is taken from the Particle Data Group (PDG)~\cite{pdg} to be $0.1181 \pm 0.002$. We 
take the best fit values for $\alpha_{em}(M_Z)$/$\sin\theta_W$ from the PDG. In principle, one should extract the $Z$-pole couplings assuming the full 
MSSM4, thereby accounting for the fourth-generation fermions and light SUSY spectra in a fully consistent way\footnote{We account for the one-loop 
leading logarithmic corrections from the SUSY sector when running the RGE by employing the step-function approach~\cite{boer}. This procedure is 
accurate in the limit of heavy sparticles but fails for masses $\tilde{M} < M_Z$ where both logarithmic and finite corrections will influence the 
extraction of the couplings.}. 
We have also performed our study in the $\overline{MS}$ scheme although it is the $\overline{DR}$ scheme that is consistent with supersymmetry. However, 
differences between the $\overline{DR}$ and $\overline{MS}$ schemes are not significant at the low-energy scale as compared to other 
uncertainties~\cite{gunion}. \medskip \\ 
In our analysis we neglect all Yukawa couplings from the first three generations, except that of the $t$-quark whose mass we take to be 
$M_t =$ 170 GeV. As is typical with four-generation models, we require small values of $\tan\beta$ (the ratio of the Higgs vev's)
so as to avoid $h_B (M_Z)\ge$$\cal O$($\sqrt{4\pi}$)~\cite{gunion}. Once all couplings at $M_Z$ have been fixed, we integrate up in energy scale using the 
two-loop renormalization group equations (RGE). The one-loop leading logarithmic threshold corrections from the SUSY sector are accounted for in the 
numerical procedure. We select the point where $\alpha_1(\mu) = \alpha_2(\mu)$ as the unification scale $M_U$ with coupling $\alpha_U(M_U)$. Any deviation 
in $\alpha_3(M_U) = \alpha_U(M_U)$, which we parameterize as $\delta = \frac{\alpha_3(M_U) - \alpha_U(M_U)}{\alpha_U(M_U)} $, can arise from either of two 
sectors. On the one hand, we have the $\overline{MS}$ vs.~$\overline{DR}$ mismatch, experimental errors in  $\alpha_{em}(M_Z)$/$\sin\theta_W$ and the 
variations in the best fit values of $\alpha_{em}(M_Z)$ and $\sin\theta_W$ as the fourth-generation and light SUSY particles ($\tilde{M} < M_Z$) are 
included. However, more importantly, assuming no intermediate scales, high-energy threshold corrections from specific GUT/string models can provide 
contributions to $\delta$. Following~\cite{ross}, we note that these corrections (for particular models) can be large. We conservatively assume that
unification is consistent if $|\delta| \le$ 5 \%. \medskip \\
The SUSY threshold corrections in the MSSM4 are important as they can influence whether or not a particular set of parameters 
($M_t,M_T,M_B,M_N,M_E,\tan\beta$) will retain perturbative consistency to the GUT scale. This can be observed analytically if we write the one-loop 
leading logarithmic correction to the strong coupling from the SUSY sector
\begin{equation}  \frac{1}{\alpha_3^+(M_Z)} - \frac{1}{\alpha_3^-(M_Z)} = \frac{b_3^{MSSM4}-b_3^{SM4}}{2\pi} \ln \left(
	\frac{M_{col}}{M_Z} \right)
\label{match}
\end{equation}
Here $M_{col}$ represents an effective threshold scale and $b_3^{MSSM4/SM4}$ represents
the one-loop beta function contribution to the strong coupling in the MSSM4/SM4 respectively. This correction is implemented at the scale $M_Z$
and accounts for the coloured sparticles with masses $M_{col} > M_Z$ in the running of the strong coupling. $\alpha^+$/$\alpha^-$ represents
the renormalized gauge coupling just above/below the scale $M_Z$ and $\alpha_3^-(M_Z)$ is fixed to be 0.118 from experimental input.  
From Eq.~(\ref{match}) we see that the higher the scale $M_{col}$, the lower $\alpha_3^+(M_Z)$. Writing the one-loop RGE for the third-generation 
top Yukawa coupling $h_t$, accounting for the $t,T,B$ quarks and the strong coupling, we have
\begin{eqnarray}
	\frac{1}{y_t} \frac{dy_t}{d\Lambda} & = & \sum_{\delta=t,T,B} a_{t\delta}y_{\delta}-\frac{16}{3}\alpha_3 \label{rgen}\\ 
	a_{t\delta} & = & (6,3,1) \nonumber
\end{eqnarray}
where $y_{\delta} = \frac{h_{\delta}^2}{4\pi}$ and $\Lambda = \frac{\ln\mu}{2\pi}$. We can see that the corrections from the 
sparticle sector (see Eq.~(\ref{match})) are constrained, since we require a large $\alpha_3^+(M_Z)$ in order to compensate for the contributions from 
$h_{T,B}$ in Eq.~(\ref{rgen}) that are driving $h_t$ to non-perturbative values. In particular, the Yukawa couplings $h_{t,T,B}$ are fixed at the 
low-energy scale through the relations
\begin{equation}
	h_{t,T} (m_{t,T}) = \frac{m_{t,T} (m_{t,T})}{\upsilon\sin\beta} \qquad
	h_B (m_B) = \frac{m_B (m_B)}{\upsilon\cos\beta}
\end{equation}
where $\upsilon =$ 174 GeV is the Higgs vacuum expectation value and $m_{t,T,B}(m_{t,T,B})$ are the quark running masses at the scale $m_{t,T,B}$. For a 
given $\tan\beta$, if $M_{t,T,B}$ is sufficiently large so that the RG running of $h_t$ is tending towards
non-perturbative values, then we obtain an upper bound $M_{col}^{max}$, above which the initial $\alpha_3^+(M_Z)$ is 
too small to counteract the effect of the fourth-generation couplings when solving the RGE in Eq.~(\ref{rgen}). \medskip \\ 
In practice, we perform a full numerical study, accounting for the threshold corrections from $t,T,B$ and all sparticles by changing the 
$\beta$-functions and using the step-function approach in the running of the gauge couplings~\cite{boer}. We assume 
separate degeneracies amongst the coloured and weak SUSY spectrum
\begin{eqnarray}
 M_{wk} & = & M_{\tilde{L}} = M_{\tilde{l}} = M_{\tilde{H}} = M_{\tilde{W}}= M_H \\
 M_{col} & = & M_{\tilde{Q}} = M_{\tilde{q}} = M_{\tilde{g}} 
\end{eqnarray}
where $l$ and $q$ denote the leptons and quarks of the first three-generations respectively.
We assume 100 GeV $\le M_{col} \le$ 2 TeV and we restrict the values of $M_{wk}$ to be 500 GeV, 1 TeV or 2 TeV. Of course, considering separate 
degeneracies amongst the coloured and weak sparticle spectra is an approximation. In fact, our model demands the introduction of light sparticles to 
ensure consistency with experimental searches for the fourth-generation and precision data bounds, thus providing significant deviations from 
degeneracy. Nevertheless, looking at Figure~\ref{susyr}, we can clearly see that the effective mass ($M_{col}$) of the coloured spectrum is bounded from 
above, in the process of retaining perturbative consistency. The effect of $M_{wk}$ is clearly weak since $\alpha_1(M_Z)$ and $\alpha_2(M_Z)$ are an order 
of magnitude lower than the strong coupling $\alpha_3(M_Z)$. In particular, although we expect the sparticles to deviate in mass around the
effective scales $M_{col}$ and $M_{wk}$, we can still conclude that as long as all the sparticles have masses less than 2 TeV, then 
perturbative unification is consistent with $(M_t,M_Q,M_L)=$ (170 GeV, 106 GeV, 46 GeV) for $1.63 \le \tan\beta \le 1.67$. \medskip \\ 
\medskip \\
\section{ The Higgs Sector of the MSSM4}
\label{sec5}
In this section, the effect of the fourth-generation on the lightest Higgs mass ($M_h$) is investigated. We shall employ the one-loop effective potential 
with contributions from the top / stop and fourth-generation fermions / sfermions. Since we are at low $\tan\beta$, we can ignore the contributions from 
the third generation bottom / sbottom masses. At tree-level, the Higgs sector of minimal supersymmetry models is fixed by two parameters $(i) \tan\beta$ 
and $(ii) M_A$ (the pseudoscalar Higgs-boson mass). We set $\tan\beta =$ 1.64 from the requirement of perturbative unification (see Figure~\ref{susyr}). 
Assuming the pseudoscalar Higgs-boson mass ($M_A$) is degenerate with the heavy Higgs sector ($M_H$) we set $M_A =$ 500 GeV. Accounting for 
one-loop terms, the Higgs mass $M_h$ becomes dependent on ($M_{t,\tilde{t}}$; $M_{T,\tilde{T}}$; $M_{B,\tilde{B}}$; $M_{N,\tilde{N}}$; $M_{E,\tilde{E}}$). 
In general, the sparticle mass matrix squared in the L/R basis can be written as
\begin{equation}
	M^2_{\widetilde{i,j}} = \left(\begin{array}{cc}
		M_{S_{i,j}}^2 + M_{i,j}^2+\Delta_{D_{i,j}} & M_{i,j} \left(A_{i,j} + \mu \cot \beta \right) \\
		M_{i,j} \left(A_{i,j} + \mu \cot \beta \right)  & M_{{\overline{S}}_{i,j}}^2 + M_{i,j}^2+\Delta_{{\overline{D}}_{i,j}} 
	\end{array} \right)
\label{mat}
\end{equation}
where $\Delta_{D_{i,j}}$, $\Delta_{{\overline{D}}_{i,j}}$ represents the $D$-term contributions; $i = t, T, B$ and $j = E, N$. Since large values 
of $|\mu|$ are required by radiative electroweak breaking and also to preserve the hierarchy in Eq.~(\ref{hier}), we take $|\mu| =$ 500 GeV. The 
sparticle mass eigenvalues squared, denoted by ${(M_{\widetilde{i,j}})}^2_{1,2}$, are obtained by diagonalizing the matrix in Eq.~(\ref{mat}). Since the 
Higgs mass is dependent on the mixing in the sparticle sector, it is insufficient to consider a fully degenerate SUSY spectrum 
as in the approximation used in section~(\ref{sec4}). To account for this, we set the soft supersymmetric breaking terms for the coloured squarks 
to be $M_{S_i} = M_{{\overline{S}}_i} = M^{soft}_{col}$ which we vary over the range 100 GeV $\le M^{soft}_{col} \le$ 1 TeV. For the sleptons we take
$M_{S_j} = M_{{\overline{S}}_j} = M^{soft}_{wk} =$ 500 GeV. We then randomly vary the two mixing 
parameters $A_i (= A_{t,T,B})$; $A_j (= A_{E,N})$ and retain the maximum value returned for the lightest Higgs mass $M_h^{max}$. As shown in previous 
studies of the Higgs sector, this will occur for $X_{i,j} \sim \pm \sqrt{6} M_{S_{i,j}}$ where $X_{i,j} = A_{i,j} + \mu\cot\beta$. Large mixings will induce 
light sparticles ($< M_Z$) which are required in the MSSM4 to provide the decay channels to the fourth-generation quarks. The minimum Higgs mass is 
obtained when $X_{i,j} \sim$ 0. \medskip \\
In the MSSM4, for $M_h \ge$ 100 GeV, the channel ($h \rightarrow \mbox{invisible}$), where invisible represents ($N\overline{N},E\overline{E},\tilde{N}
\overline{\tilde{N}}, \tilde{\chi}\overline{\tilde{\chi}}$), will open and would dominate over conventional $h \rightarrow b\overline{b}$ rates.
Exclusion limits will now come from the missing energy search ($e^+e^- \rightarrow Zh \rightarrow Z + \mbox{missing energy}$) that currently sets the 
lower bound at 114.4 GeV~\cite{invisible} assuming a SM-like Higgs and $BR(h \rightarrow \mbox{invisible}) =$ 1. We therefore need to check, for our 
constrained set of parameters ($M_{t,\tilde{t}}$; $M_{T,\tilde{T}}$; $M_{B,\tilde{B}}$; $M_{N,\tilde{N}}$; $M_{E,\tilde{E}}$; $\tan\beta$), that 
$M_h^{max}$ lies above this experimental lower bound. \medskip \\
We find a large region 370 GeV $\le M^{soft}_{col} \le$ 1000 GeV where $M_h^{max} >$ 114.4 GeV, although some mixing in the third-generation stop and/or 
fourth-generation squark sector is required. The region where $\tan\beta \sim 1.64$, although excluded in the 
MSSM3 scenario from experimental lower limits~\cite{hollik}, is allowed in the MSSM4 due to the extra one-loop contributions that arise from 
$T,\tilde{T}$; $B,\tilde{B}$; $E,\tilde{E}$ and $N,\tilde{N}$. The lightest Higgs-boson would then decay invisibly with an upper bound on its mass of
$M_h \le$ 140 GeV
\footnote{We have not accounted for two-loop terms that have been shown to correct the Higgs-mass $M_h$ by up to -10 GeV in the MSSM3~\cite{twoloop}.}. 
\section{Conclusions}
\label{sec6}
We have seen that it is possible to incorporate a fourth-generation into a GUT model, requiring the existence of supersymmetric particles in order
to provide the necessary cancellations in precision fits. The fourth-generation masses are then tightly constrained with typical values of 
$M_Q =$ 106 GeV; $M_L =$ 46 GeV. 
To retain perturbative consistency to the unification scale, assuming all sparticles have masses less than 2 TeV, we constrain $\tan\beta$ to lie in the 
interval 1.63 $\le \tan\beta \le$ 1.67. We find $M^{soft}_{col} \ge$ 370 GeV \footnote{$M^{soft}_{col}$ represents the scale of the soft breaking 
parameters in the $\tilde{t}, \tilde{T}, \tilde{B}$ sector.} from the requirement that the lightest Higgs mass must be above its present experimental 
lower bound, although some mixing between the L/R weak eigenstates in the third and/or fourth-generation squark sector is required. In order to provide 
decay channels to the fourth-generation quarks, it might be that the LSP is the sneutrino ${\tilde{N}}_1$ with a mass $M_{{\tilde{N}}_1} \sim$ 40 GeV. 
This has implications for dark matter constraints. We would also expect a light ${\tilde{B}}_1$ with $M_{{\tilde{B}}_1} \sim$ 40 - 50 GeV along with a 
degenerate neutralino/chargino pair $M_{\tilde{\chi}} =$ 55 - 65 GeV. The SUSY spectrum needed to 
satisfy all these constraints cannot be obtained from MSUGRA scenarios with universal parameters at the unification scale.\medskip \\
\section*{Acknowledgements}
We would like to thank David Sutherland for discussions and useful remarks.

\end{document}